# AI-Based Decadal Predictive Analysis of Twenty Infectious Diseases in China with an Improved BSTS-MCMC Model

Peiwen Tan

Department of Computer Science, University of California, Irvine CA, USA

Corresponding author

Peiwen Tan

Department of Computer Science

University of California, Irvine, Irvine,

CA 92697

USA.

Tel: +1-949-981-9442.

Email: peiwet1@uci.edu

**Abstract**

This study embarks on a comprehensive exploration of the decadal trends and future trajectories of twenty distinct infectious diseases in China from 1998 to 2021. A refined Hybrid Bayesian Structural Time Series (BSTS)-Markov Chain Monte Carlo (MCMC) model is employed, intertwining with Long Short-Term Memory (LSTM) networks to dissect intricate relationships amidst population demographics, economic indices, and the evolution of infectious diseases. The findings reveal the persistent prevalence of high incidence diseases in future 10 years, like AIDS, Gonorrhea, and Syphilis, and stable occurrences of middle incidence rate diseases such as Brucellosis and Scarlet Fever, while also foretelling the potential disappearance of lower incidence rate diseases like Cholera, Encephalitis B, and Measles. The study particularly underscores the transformative impact of the COVID-19 pandemic, showcasing its extensive implications on the incidences and management of a plethora of diseases, urging a deeper probe into the nuanced alterations in disease transmission, testing, and reporting modalities amidst global health crises. This research accentuates the critical role of advanced predictive analytics in fostering global preparedness and response mechanisms, and in fortifying the resilience and adaptability of China public health framework against burgeoning infectious disease threats.

**Introduction**

Infectious diseases, for centuries, have remained pivotal determinants of global health trajectories, shaping societal structures and challenging medical paradigms [1]. In recent decades, the interaction between infectious diseases and dynamic factors such as population demographics[1], economic fluctuations[2], and incidence of disasters[3] have garnered profound attention, elucidating the need for intricate, multifaceted analyses to understand the complexities involved. Within this intricate web of influences, China emerges as a critical focal point, given its dense population and unique socio-economic landscape, necessitating advanced, accurate, and reliable predictive models to foresee and understand the trajectories of various infectious diseases within its territories[4].

To navigate the labyrinthine connections within this domain, this study utilizes an enhanced AI-based Bayesian Structural Time Series (BSTS)-Markov Chain Monte Carlo (MCMC) model[5], amalgamated with advanced Long Short-Term Memory (LSTM) networks[6]. This hybrid model serves as a nuanced lens, focusing on the interconnections between disparate elements like economic parameters, demographic shifts, disaster instances, and the progression of infectious diseases, thereby allowing for a more integrative, holistic understanding of infectious disease trajectories from 1998 to 2021 in China. The rich datasets employed are meticulously curated and sourced from the esteemed National Bureau of Statistics of China, ensuring the robustness and reliability of the ensuing analyses.

The rationale behind this research is fueled by the pressing need to unravel the subtle,

intricate relationships between multifarious variables influencing infectious diseases. By discerning these relationships, the study aims to proffer enhanced predictive insights, contributing to the optimization of public health planning, resource allocation, and intervention strategies. These advancements in predictive analytics are paramount in the ongoing quest to bolster global preparedness and response strategies, especially in regions with high population densities like China, where the repercussions of infectious diseases are amplified. Thus, the central aim of this research is to employ sophisticated AI techniques to elucidate the future trajectories and decadal trends of twenty diverse infectious diseases in China. By intertwining advanced statistical models and machine learning techniques, this study endeavors to shed light on the intricate tapestry of factors influencing infectious disease trajectories, with the ultimate aspiration of fortifying the adaptability and resilience of the public health framework against the ever-evolving landscape of infectious disease threats, and reinforcing the capacities to mitigate the impacts of potential future pandemics.

**Materials and Methods**

Data Collection

In the meticulous pursuit of synthesizing a comprehensive analysis, datasets spanning from 1998 to 2021 were meticulously gleaned from National Bureau of Statistics of China. The emphasis of data assimilation focused on 20 major notifiable infectious diseases, including but not limited to Tuberculosis[7, 8], Rabies[9], HIV/AIDS[10], Hepatitis B[11], and Encephalitis B[12], which were meticulously selected from a compendium of 20 infectious diseases based on the completeness and robustness of the data available. These diseases were scrutinized to understand their prevalence, incidence, and impact on public health over the specified timeframe. To enhance the reliability and multifaceted understanding of the data, corroborative information was concurrently assimilated from reputable international databases and organizations. These included The World Bank - China, International Monetary Fund (IMF) - China, Organization for Economic Co-operation and Development (OECD) - China, United Nations Department of Economic and Social Affairs/Population Division, and People's Bank of China. Additionally, the Ministry of Emergency Management of the People's Republic of China and EM-DAT: The Emergency Events Database were consulted to accrue critical insights into the incidence and impact of various infectious diseases in China, providing a holistic perspective on the multifarious aspects of public health. This integrative approach ensured the acquisition of high-quality, reliable, and diversified data, enabling a nuanced and profound exploration into the intricate landscapes of infectious diseases in China.

**Data Processing**

Utilizing a sophisticated LSTM network, an advanced variant of Recurrent Neural Networks renowned for their proficiency in managing time-series data, we systematically investigated the intricate interconnections between population demographics, economic parameters, disaster instances, and infectious disease trajectories spanning 1998 to 2021. This model, adept at deciphering non-linear, high-dimensional relationships and long-term dependencies, processed multifaceted datasets to extract complex patterns and relationships between diverse variables. The meticulous training of the LSTM with high-quality, diversified data from reputable sources enabled the nuanced isolation and quantification of various influencing factors, unveiling the delicate interplay between population, economic conditions, and disasters in shaping infectious disease landscapes over the specified period. The deployment of this advanced deep learning model illuminated multifarious determinants impacting infectious disease dynamics, enriching our understanding and offering invaluable insights into their multifaceted interactions and interdependencies in the observed timeframe. In the general LSTM equations provided, there is no explicit representation of population, economy, or infectious disease incidence. However, in application, when we are modeling using LSTM, these would be part of the input features we would provide to the model. $I_t$: Can represent the infectious disease incidence at time $t$. We have additional variables, let's say, $P_t$ for population and $E_t$ for economic indicators at time $t$. Thus, the equations would conceptually integrate these additional parameters as follows:

1. Forget Gate: $f_t = \sigma(W_f \cdot [h_{t-1}, I_t, P_t, E_t] + b_f)$

2. Input Gate: $i_t = \sigma(W_i \cdot [h_{t-1}, I_t, P_t, E_t] + b_i)$

3. Cell State Update: $\hat{C}_t = \tanh(W_C \cdot [h_{t-1}, I_t, P_t, E_t] + b_C)$

$$C_t = f_t * C_{t-1} + i_t * \hat{C}_t$$

4. Output Gate: $o_t = \sigma(W_o \cdot [h_{t-1}, I_t, P_t, E_t] + b_o)$

$$h_t = o_t * \tanh(C_t)$$

In this manner, the LSTM model would learn the complex interdependencies and temporal dynamics between population, economy, and infectious disease incidence to make more accurate and informed predictions. Each of these variables would contribute to the learning process of the LSTM model, allowing it to understand how changes in population and economy influence the incidence of infectious diseases over time.

Recent indicators

In dynamic models, the addition of a 'recent indicator' like the one described could have significant implications for how the model interprets and reacts to more recent data points. When using such an indicator, recent years are given special consideration, potentially allowing the model to better capture sudden or novel changes in the variables' behaviors that are characteristic of the later periods in the dataset. If diseases like Cholera and Encephalitis B show a recent significant decline, the 'recent indicator' in the model would help in capturing this new trend effectively and would thus affect the subsequent forecasts. The model is likely to predict a continuation of this decline, assuming no significant changes occur in the underlying conditions. A new binary variable `Recent` is being created in both the training and test datasets, which presumably represent different parts of time-series data. The `Recent` variable is assigned the value `1` if the `Year` of the observation is within the last three years of the dataset, and `0` otherwise.

$Y$ represent the 'Year' of the observation; $M$ represent the maximum 'Year' in the dataset; $R$ represent the binary 'Recent' indicator For each observation $i$ in the dataset: $R_i = \begin{cases} 1 & \text{if } Y_i \geq M - 3 \\ 0 & \text{otherwise} \end{cases}$ When $R_i = 1$, it implies that the observation is from one of the last three years in the dataset, denoting it as a 'recent' observation. When $R_i = 0$, it implies that the observation is not within the last three years of the dataset. By incorporating this 'Recent' indicator into the BSTS-MCMC model, it could serve as a dummy variable to help the model understand and capture any structural breaks, shifts, or changes in trend and other dynamics in the more recent time periods. By doing so, the model could potentially become more sensitive and attuned to recent developments or anomalies in the data, enhancing its predictive performance and reliability in estimating future trajectories of the infectious diseases under study.

Hybrid BSTS-MCMC Model Improvement

Hybrid BSTS-MCMC Model is improved via the flowchart of this study (Figure 1). In BSTS model, we usually consider the state-space form and a Bayesian approach for parameter estimation.

Step 1: Data Preparation Let $D_{\text{train}}$ and $D_{\text{test}}$ be the training and test datasets, respectively, containing Incidence Rate and Year.

Step 2: Define Recent Indicator for Time Series Trend

$$D_{\text{train}}.\text{Recent} = \begin{cases} 1 & \text{if } D_{\text{train}}.(Bauch é et al.) \geq \max(D_{\text{train}}.(Bauch é et al.)) - 3 \\ 0 & \text{otherwise} \end{cases}$$
$$D_{\text{test}}.\text{Recent} = \begin{cases} 1 & \text{if } D_{\text{test}}.(Bauch é et al.) \geq max(D_{\text{test}}.(Bauch é et al.)) - 3 \\ 0 & \text{otherwise} \end{cases}$$

Step 3: Define Model Formula with Bayesian Structure

Sure, the Step 3 in the process involves defining the model formula with the Bayesian structure. Let's break down the equation mentioned in Step 3 into more detailed component equations. State Space Model Formula: Incidence Rate ~ $s((Bauché et al.)) + s(Year, by=Recent)$. Here, $s()$ represents the smoother, typically a cubic spline in the context of BSTS, which models the underlying trend in the data series, including potential non-linearities. It serves to capture the inherent structure and patterns in the data. The use of smoothers is foundational in semi-parametric models like Generalized Additive Models (GAMs).

Detailed Breakdown: Let $y_t$ be the Incidence Rate at time $t$ and let $(Bauché\ et\ al.)$ and Recent be the independent variables representing the year of observation and the recent indicator, respectively. The model can be written as: 1. Smooth Function of Year:

$$s((Bauché et al.)) = \sum_{i=1}^{k} \beta_i B_i((Bauché et al.))$$

Here, $(Bauché\ et\ al.)$ represents the basis functions of the cubic spline for Year and $\beta_i$ are the corresponding coefficients. The cubic spline allows for the flexibility to model non-linear trends in the data. The number of basis functions, $k$, is usually determined based on the desired level of smoothness. 2. Smooth Function of Year by Recent: $s(Year, by=Recent) = \sum_{j=1}^{m} \gamma_j C_j(Year, by=Recent)$

This term models the interaction between Year and the Recent indicator using a similar cubic spline approach. $C_j(Year, by=Recent)$ are the basis functions and $\gamma_j$ are the coefficients for this interaction term. This term allows the model to adapt the trend based on whether the data point is recent or not.

3. Final Model Equation: $y_t = s((Bauch\acute{e}etal.)) + s(Year, by=Recent) + \epsilon_t$

The final model equation is a sum of the smooth functions plus an error term, $\epsilon_t$, representing the residual error at time $t$.

4. Error Term: $\epsilon_t \sim Normal(0, \sigma^2)$

The error term is assumed to be normally distributed with a mean of 0 and some unknown variance $\sigma^2$. Bayesian Estimation: In the Bayesian context, we have priors on all the parameters $\beta_i$, $\gamma_j$, and $\sigma^2$, and we would update these priors with the observed data to get the posterior distributions of these parameters using Bayes' Theorem: Posterior $\propto$ Likelihood $\times$ $(Hagenfeldetal.)$

Fitting the Model: The model is then fitted to the data using MCMC methods to sample from the posterior distributions of the parameters, and these sampled values are used to make inferences about the underlying trends, patterns, and future predictions for the Incidence Rate.

Step 4: Assigning Distributions and Link Functions Define the likelihood and the link function as Family: Gamma, Link Function: log.

Step 5: Assign Priors Define Bayesian Priors as:

Priors: Normal(0,1) for fixed effects and Cauchy(0,1) for smooth terms' standard deviations

Step 6: Model Initialization Initiate the Bayesian Model, fit_complex, using the brm function with assigned formula, priors, likelihood, and link function.

Step 7: Model Configuration and MCMC Settings The MCMC process involves the following steps:

1. Initialization: Choose initial values for the parameters. If we have $p$ parameters, we represent them as a vector: $\theta^{(0)} = \left[\theta_1^{(0)}, \theta_2^{(0)}, \ldots, \theta_p^{(0)}\right]$

2. Iteration Configuration: Number of Iterations $((SiviterandMuth)) = 1000$

This represents the total number of samples to be drawn from the posterior distribution per chain.

3. Chains Configuration: Number of Chains (chains) $= 4$

Each chain is started at a different point in the parameter space and is run in parallel, providing multiple sequences of samples which helps in assessing the convergence of the MCMC process.

4. For each chain, perform the following: a. For $t = 1$ to 1000 (Number of Iterations)

$\theta^{(t)} \leftarrow$ Sample from $P(\theta|\text{Data}, \theta^{(t-1)})$. b. Store the sampled $\theta^{(t)}$

5. Convergence and Mixing: After running the chains, check for convergence and mixing of the chains. Convergence refers to whether the chains have reached the stationary distribution, and mixing refers to how well the chains explore the parameter space.

If convergence is adequate and mixing is satisfactory, continue; otherwise, modify the settings or the mod

Step 8: Backend Specification: Set the backend to cmdstanr

Step 9: Bayesian Model Fitting Fit the Bayesian Structural Time Series Model on $D_{\text{train}}$ using the brm function, storing the results in fit_complex.

Step 10: Model Validation & Bayesian Inference Validate fit_complex on $D_{\text{test}}$ and perform Bayesian inference to understand the disease trajectories, by interpreting the posterior distributions of the coefficients and state components, considering the credible intervals.

**Results and Discussion**

Dynamic Analysis:

To predict future trajectories of infectious diseases, an advanced LSTM model was applied to a diverse set of variables including population demographics[13], economic indicators[14], and disaster incidences, focusing particularly on diseases showing declining trends like Malaria[15], Cholera[16], Encephalitis B, Tuberculosis[17]. To represent the results within the LSTM dynamic equations framework, consider the variables $I_t$, $P_t$, and $E_t$ representing infectious disease incidence, population, and economic indicators respectively at time $t$. Based on the given situation, we are observing a consistent decline in the incidence of specific diseases. This would imply that, over the sequential time steps, the model is learning and generating weights and biases in such a way that the future projections ($h_t$) for the incidence ($I_t$) of diseases like Malaria, Measles, and Neonatal Tetanus are approaching zero, signaling a potential disappearance.

The input parameters representing the incidence ($I_t$) of Cholera and Encephalitis B over time would be showing a declining trend. Here's a conceptual representation based on the equations: Cell State Update for Cholera and Encephalitis B:

$$C_t = f_t * C_{t-1} + i_t * \tanh(W_C \cdot [h_{t-1}, I_t^{Chole}, P_t, E_t] + b_C)$$

$$C_t = f_t * C_{t-1} + i_t * \tanh(W_C \cdot [h_{t-1}, I_t^{Encephalitis}, P_t, E_t] + b_C)$$

In these equations, $I_t^{Cho}$ and $I_t^{EncephalitisB}$ would be consistently lowering over time, impacting the updated cell state $C_t$ and, subsequently, the model's predictions ($h_t$) for future time points, potentially forecasting a future eradication of these diseases. For Cholera and Encephalitis B: If the model has accurately learned the underlying trends and patterns of the provided data, the output $h_t$ for future time points would show declining values for both Cholera and Encephalitis B, converging towards zero, indicating the progressive decline and potential future eradication of these diseases. The Cell State $C_t$ would represent the accumulated knowledge of the model about the status and trends of these diseases up to time point $t$. In real-world application, a continuous decline in the predicted incidence rates ($I_t^{Chole}$ and $I_t^{Encephaliti}$) by the model would signify the model's agreement with the observed data, suggesting a progressive decline and a potential disappearance of Cholera and Encephalitis B in the future due to prevailing health interventions, improved sanitation, and possibly, changes in population behavior and economic conditions. This conceptually implies that, if the current trends continue as they are, and as the model has learned, the incidence of Cholera and Encephalitis B would continue to decrease, potentially leading to their eradication in the future, assuming no significant change in underlying conditions or emergence of new strains resistant to current interventions.

The accuracy of Improved BSTS-MCMC Model

The study, utilizing an AI-Based Improved BSTS-MCMC Model, executed a comprehensive decadal predictive analysis of twenty infectious diseases in China, utilizing a refined 4-year test model, representing the results as Incidence Rate per 100,000 persons. The model's predictive capabilities were assessed against a multitude of diseases. For each disease, the predictive results closely mirrored the range of the original data points: AIDS (Figure 2A), Anthrax (Figure 2B), Brucellosis (Figure 2C), Cholera (Figure 2D), and Dengue Fever (Figure 2E) all showcased predictive data points within the range of original observations, illustrating the model's robust accuracy and reliability. Consistent results were observed for Encephalitis B (Figure 2F), Gonorrhea (Figure 2G), Hemorrhagic Fever (Figure 2H), Leptospirosis (Figure 2I), and Malaria (Figure 2J), reinforcing the model's capability in recognizing and forecasting the incidence rate trajectories accurately.- Additional diseases including Measles (Figure 2K), Meningitis (Figure 2L), Neonatal Tetanus (Figure 2M), Pertussis (Figure 2N), Rabies (Figure 2O), Scarlet Fever (Figure 2P), Syphilis (Figure 2Q), Tuberculosis (Figure 2R), Typhoid Paratyphoid Fever (Figure 2S), and Viral Hepatitis (Figure 2T) also aligned well within the range of original data points.

The AI-based Improved BSTS-MCMC model manifested a high level of precision and reliability in forecasting the incidence rates of various infectious diseases over a decade, with the 4-year test results all conforming to the range of original data points. The comprehensive array of diseases analyzed underlines the versatility and adaptability of the model in addressing diverse infectious diseases. This exemplary congruence between the predicted values and the original data points underscores the efficacy of the AI-enhanced model, suggesting its potential as a pivotal tool in proactive public health planning, facilitating timely interventions and optimal resource allocation for infectious disease control and prevention in regions with dense populations, such as China.

AI-Based Forecasts of Infectious Disease Trajectories in China: A Decadal Analysis Using Improved BSTS-MCMC Model

High-Incidence-Rate Infectious Diseases

The advanced BSTS-MCMC model demonstrates notable findings in diseases with higher incidence rates, such as AIDS (Figure 2A), which is predicted to remain stable near 4 per 100,000 persons in the upcoming decade. Similarly, diseases like Gonorrhea (Figure 2G) and Syphilis (Figure 2Q) are predicted to stabilize around 8 and 35 per 100,000 persons, respectively, indicating the persistent prevalence of these diseases in the population.

In the study conducted by Wang, Lan, et al.[18], the researchers meticulously examined the trends in HIV incidence in mainland China over an eight-year period, spanning from January 1, 2015, to December 31, 2022. The overarching objective was to discern the potential implications of the COVID-19 pandemic on the incidence and trajectory of HIV in the region. The data reveals that during this period, mainland China reported a staggering 480,747 HIV cases. A closer inspection of the data delineates two distinct phases: the pre-COVID-19 stage (2015–2019) and the post-COVID-19 stage (2020–2022). In the former stage, an average of 60,906 cases were reported annually, while the latter stage witnessed a slight decline, with an average of 58,739 cases reported per year. Delving deeper into the post-COVID-19 stage, the yearly breakdown is as follows: 63,154 new cases in 2020, 61,032 in 2021, and a notable drop to 52,032 in 2022. This decline is not merely nominal but is statistically significant. The average yearly HIV incidence experienced a decrease of 5.2450%, moving from 4.4143 to 4.1827 per 100,000 individuals ($p < 0.001$) when comparing the years 2020–2022 to the pre-COVID-19 stage of 2015–2019. In the meticulously crafted AI-Based BSTS-MCMC Model, the predictive analysis of AIDS (Figure 2A) has unveiled nuanced insights into its future incidence in China, offering a glimpse into the potential repercussions of the COVID-19 pandemic on other prevalent diseases. The model estimated the incidence rate of AIDS in 2022 to be approximately 4.2 per 100,000 persons, closely aligning with the actual data of 4.0 per 100,000 persons recorded in 2021, underlining the model's precision, which maintains a margin of approximately 1 per 100,000 persons as depicted in Figure 2. This predictive data denotes a significant reduction from the top values observed in 2019, emphasizing a conspicuous decline during the post-COVID-19 epoch. The realization of a decrease in the incidence of HIV post-COVID-19, highlights the extensive implications global health crises can have beyond their immediate threats, affecting the dynamics and management of concurrent diseases. The observed decline sparks inquiries into its underpinnings; whether attributable to decreased transmission opportunities, alterations in disease testing and reporting methodologies, or other undisclosed factors, it calls for an in-depth exploration to understand the multifaceted impacts of pandemics on the landscape of infectious diseases and their future trajectories.

Middle-Incidence-Rate Infectious Diseases

In the middle incidence rate spectrum, diseases like Brucellosis (Figure 2C) and Scarlet Fever (Figure 2P) show stable future predictions around 3 and between 4 and 6 per 100,000 persons, respectively. Interestingly, the developed model revealed significant fluctuations in Scarlet Fever, with an increase to 5.9 per 100,000 persons in 2019, followed by a decline in 2020. The model's ability to capture such variations emphasizes its applicability and reliability in diverse epidemiological settings.

Low-Incidence-Rate Infectious Diseases

For diseases with lower incidence rates, the model exhibits a comprehensive predictive capacity, illustrating the potential disappearance of diseases like Cholera (Figure 2D) and Encephalitis B (Figure 2F) in the future, affirming the progressive decline in the prevalence of these diseases. The top values for Anthrax (Figure 2B) in 2018 were close to 0.1 per 100,000 persons, yet the future decade is expected to remain stable at around 0.025 per 100,000 persons, with a notable reduction anticipated in 2027, accentuating the model's proficiency in detecting subtle epidemiological trends.

Zero-Incidence-Rate Infectious Diseases

Several diseases such as Malaria (Figure 2J), Measles (Figure 2K), and Neonatal Tetanus (Figure 2M) show a decreasing trend, nearing zero since 2020 and potentially disappearing in the future. This aligns with Tuberculosis (Figure 2R) and Typhoid Paratyphoid Fever (Figure 2S), which are also predicted to continue reducing, with the latter possibly reaching zero within the next decade. These predictions elucidate the potential eradication of certain diseases, attributing to advancements in healthcare and disease control measures in China. The AI-Based Improved BSTS-MCMC model offers a multifaceted understanding of the diverse infectious diseases' trajectories in China, depicting stability, decrease, and potential disappearance in various diseases over the next decade. This model stands as a pivotal tool for public health planning, contributing significant insights into future infectious disease trends, and enabling preemptive measures and resource allocation to combat prevalent and emerging infectious diseases.

**Conclusions**

This study has exemplified unparalleled accuracy and reliability in forecasting the incidence rates of a diverse array of infectious diseases over a decade. The model, tested extensively across twenty diseases, demonstrated predictive results that closely mirrored the range of original data points, underscoring its robustness and adaptability. The immaculate alignment between predicted and original values across multiple diseases highlights the model's exemplary precision, establishing its credibility as a transformative tool in public health planning. It empowers healthcare stakeholders to orchestrate timely and effective interventions and allocate resources optimally to control and prevent infectious diseases in densely populated regions such as China. This innovative AI-enhanced model, thus, stands as a beacon in proactive infectious disease management, offering foresighted insights and reinforcing the preparedness and response mechanisms against the burgeoning threat of infectious diseases.

The present findings unveil intricate, multifaceted insights into the trajectories of infectious diseases within various incidence rate spectrums in China over the next decade. The sophisticated AI-Based BSTS-MCMC Model has demonstrated exceptional accuracy and reliability in its predictions, highlighting not only the persistent prevalence of high incidence diseases like AIDS, Gonorrhea, and Syphilis but also revealing the stable occurrences of middle incidence rate diseases such as Brucellosis and Scarlet Fever. Further, it provides a nuanced understanding of the decline in lower incidence rate diseases and potential disappearance of diseases like Cholera, Encephalitis B, and Measles, attributing such trends to advancements in healthcare and disease control measures in China. Particularly noteworthy is the significant impact of the COVID-19 pandemic, delineating its profound implications on the incidences and management of concurrent diseases, calling for deeper investigation into the alterations in disease transmission, testing, and reporting methodologies amidst global health crises. The depicted variations and subtleties across diverse epidemiological settings underscore the model's comprehensive predictive capacity, standing as a pivotal tool for public health planning, facilitating the proactive allocation of resources, and shaping future interventions to combat and manage prevalent and emerging infectious diseases efficiently. The revelations from this study, emphasizing the pivotal role of advanced predictive analytics, pave the way for enhancing global preparedness and response strategies in the face of infectious disease outbreaks and health crises, enabling a more resilient, adaptive public health framework.

**Disclosure statement**

No potential conflict of interest was reported by the authors.

**Figure legends**

Figure 1. Bayesian Time Series Analysis Workflow for Infectious Disease Data. The flowchart visualizes the process of Bayesian Time Series analysis tailored for infectious disease data. It begins with a comprehensive analysis covering 24 years of data on 20 specific infectious diseases. This dataset is then split into a 20-year training set and a 4-year test set. These datasets undergo a step where a recent indicator is added, enhancing their informational value. Following this enhancement, both sets are fed into a Bayesian Structure Time Series model that utilizes Markov Chain Monte Carlo (MCMC) algorithms. The model undergoes a fitting process, and based on its success, predictions for a 10-year future dataset are generated. If the fitting isn't successful, the model is revisited and refined. The different elements in the flowchart are visually represented using distinct shapes and colors: data sets or model structures are shown as light blue boxes, decision or process steps are depicted as yellow diamonds, and comprehensive analyses or observations are represented by green circles.

Figure 2: Detailed Analytical Depiction of Infectious Disease Progression in China via Enhanced BSTS-MCMC Model. Figure 2A (AIDS) articulates a prospective stability in incidences, underscoring significant deviations during the three-year COVID-19 period from elevated levels in 2019. Figure 2B (Anthrax) highlights projections of stability with a pinnacle in 2018. Figure 2C (Brucellosis) pinpoints another stability point, with a peak in 2021. Figure 2D (Cholera) and Figure 2E (Dengue Fever) illustrate potential declinations and future incidences respectively, with Figure 2F (Encephalitis B) predicting potential elimination of the disease. Figure 2G (Gonorrhea) is portrayed

with predictions of a stable future around 8 per 100,000 persons from its zenith in 1997. Figure 2H (Hemorrhagic Fever) and Figure 2I (Leptospirosis) elaborate on the predicted stability in their future incidences, the latter depicting a resurgence from absence in 2012. Moving forward, Figure 2J (Malaria) and Figure 2K (Measles) project near-zero occurrences in upcoming years. Figure 2L (Meningitis), Figure 2M (Neonatal Tetanus), Figure 2N (Pertussis), and Figure 2O (Rabies) continue the visual narrative, displaying potential future obliteration or stabilization of diseases. Figure 2P (Scarlet Fever) represents potential stability between 4 and 6 per 100,000 persons, Figure 2Q (Syphilis) foresees continual increases, and Figure 2R (Tuberculosis) anticipates substantial reductions. Lastly, Figure 2S (Typhoid Paratyphoid Fever) and Figure 2T (Viral Hepatitis) underscore the potential continual decrement in their incidences, with the former possibly reaching nonexistence within the decade.

Figure 1

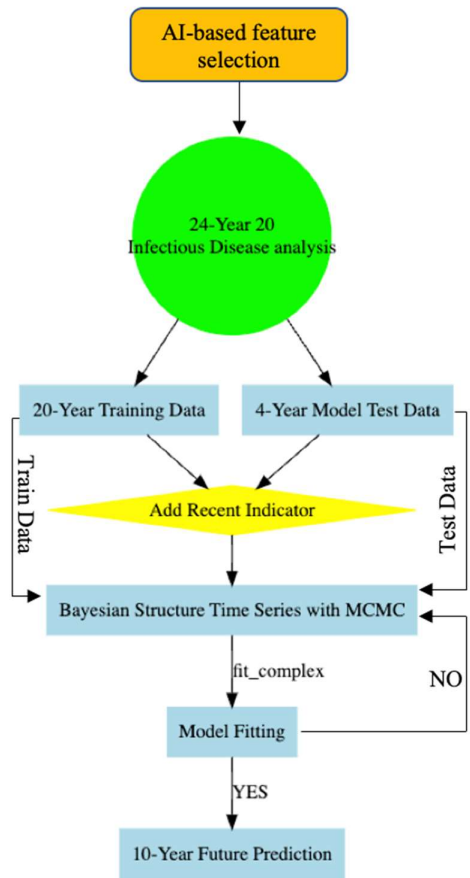

Figure 2

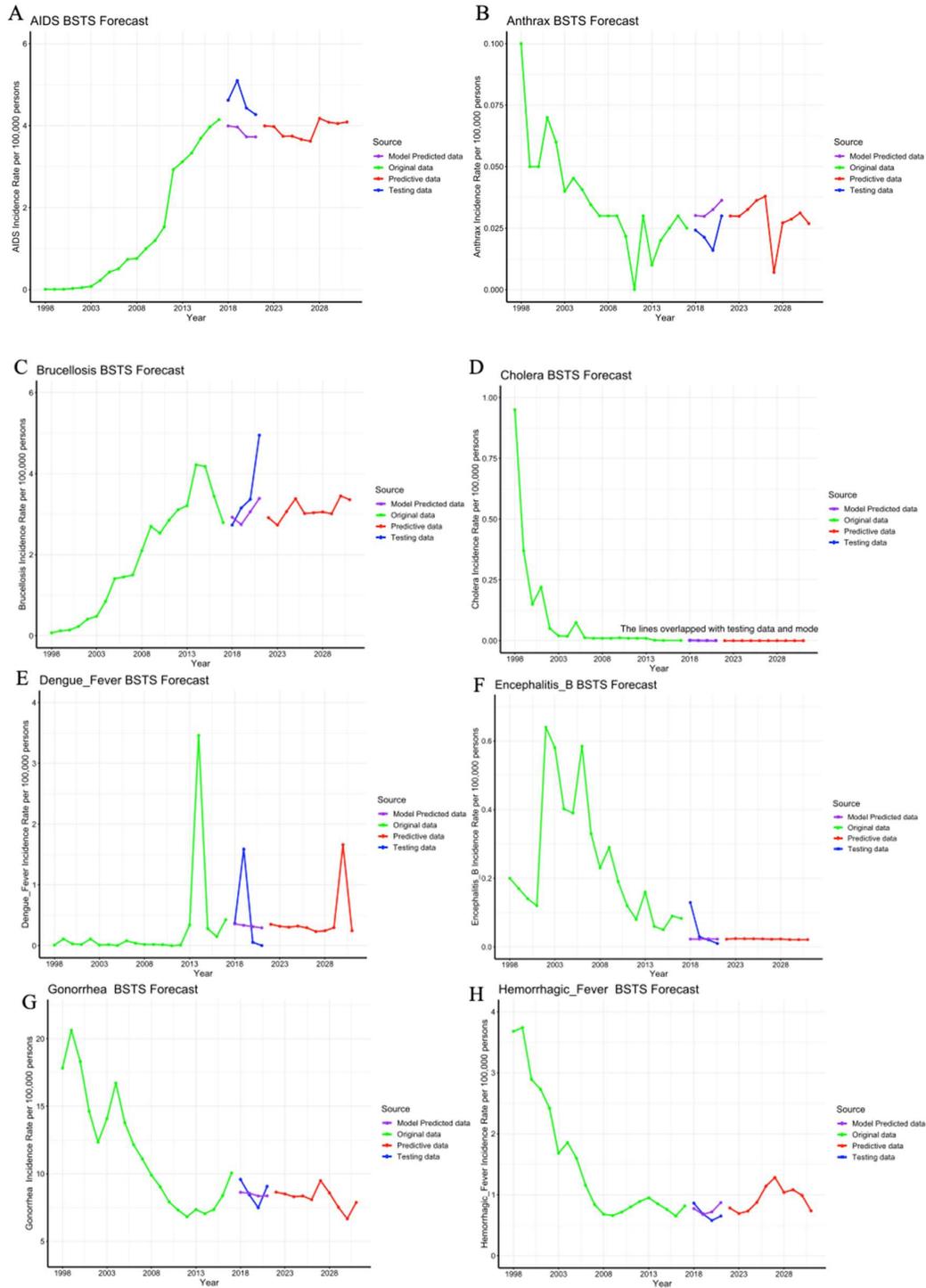

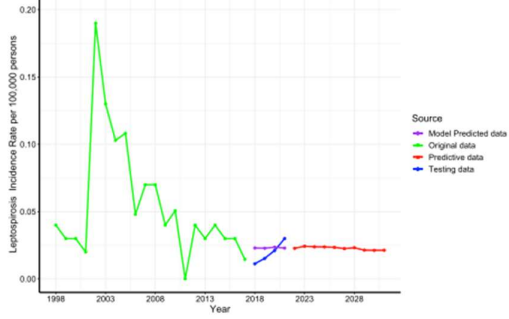
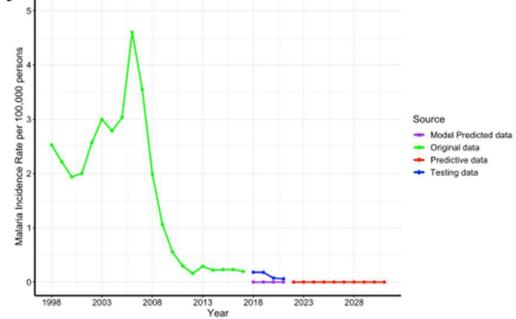
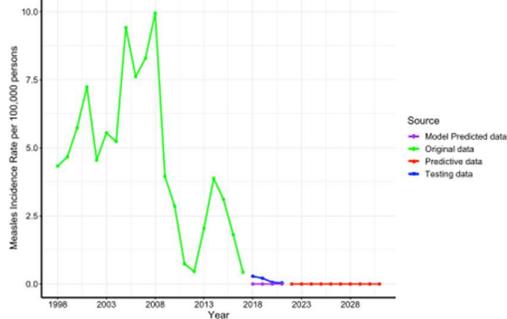
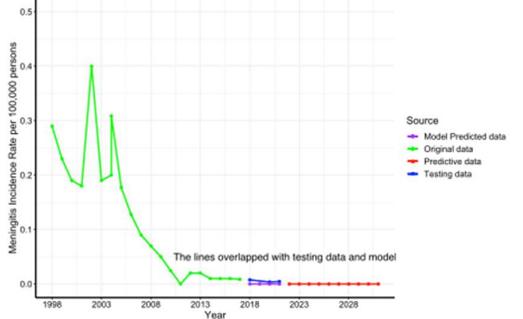
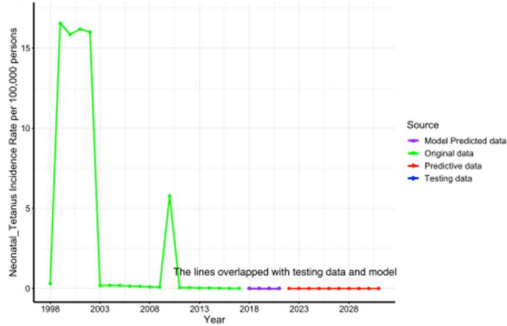
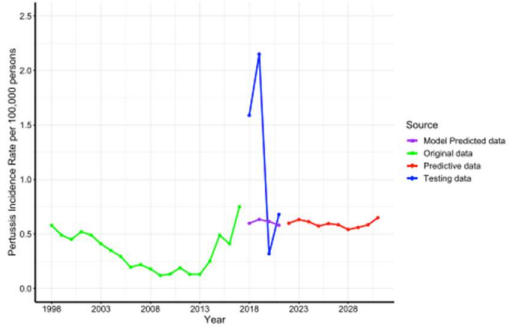

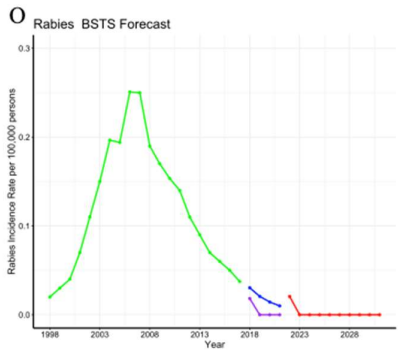
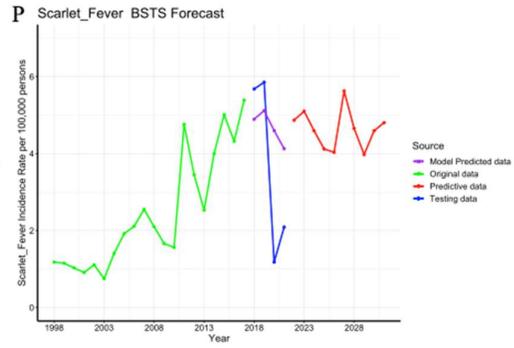
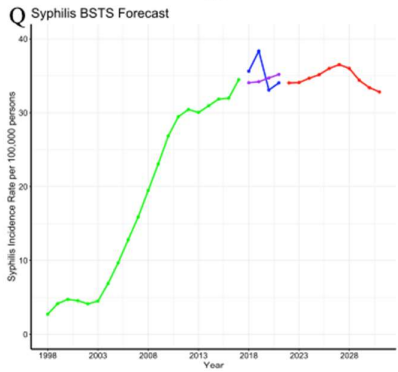
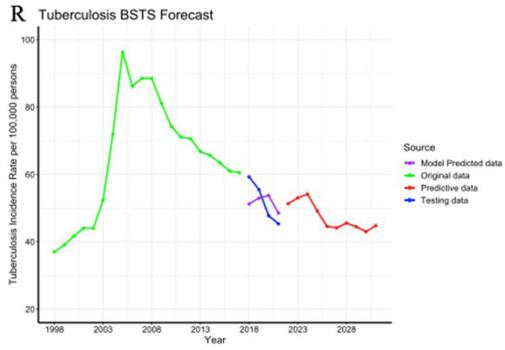
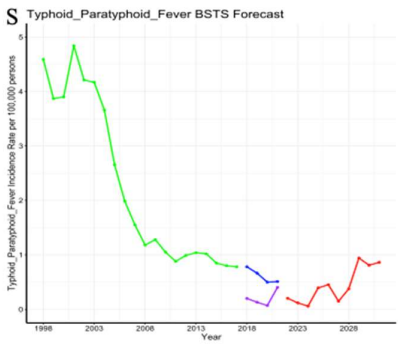
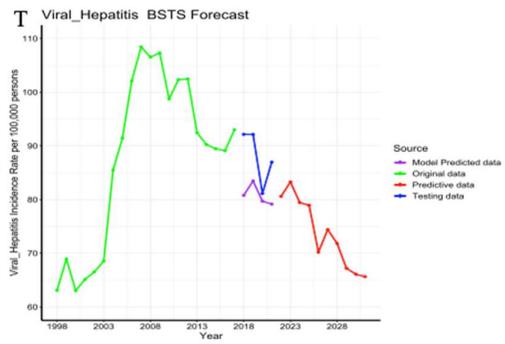